\documentclass[UTF8]{article}
\usepackage{amsmath}
\usepackage{graphicx}
\usepackage{float}

\usepackage[left=1cm,right=1cm,top=3cm,bottom=3cm]{geometry}

\usepackage{a4wide}
\usepackage{amsmath}
\usepackage[
      colorlinks=true,
      linkcolor=blue,
      urlcolor=blue,
      filecolor=black,
      citecolor=blue,
            ]{hyperref}
\usepackage{color}
\usepackage{amsfonts}
\usepackage{amssymb}
\usepackage{epstopdf}

\begin{document}

\title{\Large \textbf{On Thermodynamics of AdS Black Holes with Scalar Hair}}
\author{\large Li Li$^{\,a,b,c}$\footnote{liliphy@itp.ac.cn}~,
\\
\\
\small $^a$CAS Key Laboratory of Theoretical Physics, Institute of Theoretical Physics, \\
\small Chinese Academy of Sciences, Beijing 100190, China\\
\small $^b$School of Physical Sciences, University of Chinese Academy of Sciences, \\
\small No.19A Yuquan Road, Beijing 100049, China\\
\small $^c$School of Fundamental Physics and Mathematical Sciences, Hangzhou Institute for Advanced Study, \\\
\small UCAS, Hangzhou 310024, China
}
\date{}

\maketitle
\begin{abstract}
It has been known that in the presence of a scalar hair there would be a distinct additional contribution to the first law of black hole thermodynamics. While it has been checked in many examples, a deeper understanding of this issue is necessary. The thermodynamics of AdS black holes in Einstein-scalar gravity is studied by using the standard holographic renormalization procedure and the variation of the Hamiltonian via the Wald formalism. It is found that the first law requires a modification by including an additional term that has a particular form $\sim\left<O\right>\delta \phi_s$, with $\phi_s$ and $\left<O\right>$ a new pair of thermodynamic conjugate variables. $\phi_s$ is the leading source term of the asymptotic fall-off of the scalar field near the AdS boundary, and $\left<O\right>$ is precisely the response of the dual scalar operator from the holographic point of view. Some hairy black holes are constructed explicitly to check the first law of thermodynamics as well as the thermodynamic relations.
\end{abstract}

\newpage
\tableofcontents

\section{Introduction}
The discovery of the thermodynamics of black holes has revealed a deep and fundamental relationship among gravitation, thermodynamics and quantum theory. It was achieved primarily by using classical and semiclassical analysis and has been given rise to most of our present physical understandings into the nature of quantum phenomena in the strong gravity regime~\cite{Ross:2005sc,Wald:1999vt}. Over a long period in the past, it was established that black holes are uniquely specified by asymptotic charges, because of a number of black hole theorems. Nevertheless, more recently, motivations from higher dimensions, string theory and holography have led to considerations that violate many of the assumptions of these black hole theorems. Many new black hole solutions have been constructed and investigated, and the physics of black holes becomes more abundant than previously believed~\cite{Dias:2015nua,Chesler:2013lia}. One of the simplest and most interesting cases is black holes with a scalar field, and there are increasing number of black hole solutions with non-trivial scalar hair in the literature. Despite recent progress, the subject of scalar hair and its contribution to the first law of black hole thermodynamics remain to be elucidated.

It has been shown~\cite{Liu:2013gja,Lu:2014maa} that the naively-expected first law of thermodynamics $d E=T d S$ for the Einstein-scalar black holes does not hold, where $E$ is the energy or mass of a black hole, and $T$ and $S$ are the Hawking temperature and the Bekenstein-Hawking entropy, respectively. More precisely, for the (d+1)-dimensional Einstein-scalar gravity
\begin{eqnarray}\label{action}
S=\frac{1}{2\kappa_N^2}\int d^{d+1}x \sqrt{-g} \left[\mathcal{R}-\frac{1}{2}\nabla_\mu \phi \nabla^\mu \phi-V(\phi)\right]\,,
\label{Smatter}
\end{eqnarray}
the first law of AdS black holes should be modified with the full differential $d E$ being shifted by a 1-form:
\begin{equation}\label{Zterm}
Z\equiv-X dY=c_1\phi_v d\phi_s-c_2\phi_s d\phi_v\,,
\end{equation}
where $\phi_s$ and $\phi_v$ are two coefficients of the asymptotic fall-off of the scalar field near the AdS boundary. The first law then becomes
\begin{equation}\label{firstlaw0}
d E=T dS+X dY=T dS-(c_1\phi_v d\phi_s-c_2\phi_s d\phi_v)\,.
\end{equation}
The two constants $c_1$ and $c_2$ depend on the spacetime dimension and the mass of the scalar field. In particular, it was emphasized that $c_2\neq-c_1$, and therefore the contribution in~\eqref{Zterm} is not integrable. While it has been argued that $X$ and $Y$ describe the potential and charge of the scalar hair that breaks some of the asymptotic AdS symmetries~\cite{Lu:2013ura}, the physical meaning of $X$ and $Y$ is unclear.\,\footnote{$Y$ was considered as the scalar charge in~\cite{Lu:2013ura}, but no conserved current associated with this charge has been known so far.} As the Einstein-scalar gravity is the simplest case in a variety of gravity and supergravity theories, and has wide applications in many fields, it would be of great interest to have a further understanding on the first law of black holes with non-trivial scalar hair and the physical meaning of the new pair of variables $X$ and $Y$. In particular, in many studies of holographic QCD the black holes with scalar hair were constructed to mimic QCD phenomenology in a strongly coupled regime, for which some physical observables were obtained by using the traditional first law of black holes together with some thermodynamic relations, see \emph{e.g.}~\cite{Gubser:2008ny,Li:2011hp,He:2013qq,Rougemont:2015wca,Fang:2019lsz}.\,\footnote{For example, the pressure or free energy is computed directly by integrating the first law of thermodynamics, while keeping all sources fixed.} However, if the first law needs to be modified in the presence of the scalar hair by the form~\eqref{firstlaw0}, one must take the new modification~\eqref{Zterm} into consideration when discussing the equation of state and other thermodynamic properties. 

Although there are some discussions to deal with this issue in the literature, most of them imposed a special relationship on the coefficients $\phi_s$ and $\phi_v$ (see \emph{e.g.} ~\cite{Hertog:2004ns,Amsel:2006uf,Henneaux:2006hk,Anabalon:2015xvl}). In this so-called ``designer gravity''  the boundary conditions of the scalar in the AdS boundary are modified, such that $\phi_s$ and $\phi_v$ are independent. In some sense, it means that only a special branch of hairy black hole solutions will be considered. This raises an interesting question: Is it possible to formulate a unified version of the first law of black hole thermodynamic without imposing a special relationship on the coefficients $\phi_s$ and $\phi_v$? In the present paper, we investigate the thermodynamics of the AdS black holes in Einstein gravity coupled to a scalar field described by the action~\eqref{action}. We adopt the standard holographic renormalization procedure~\cite{Skenderis:2002wp} to define thermodynamic variables, such as energy, pressure and free energy. Then we use the Wald formalism~\cite{Wald:1993nt,Iyer:1994ys} to derive the first law of black hole thermodynamics. We show how the scalar hair contributes non-trivially to the first law of thermodynamics. In contrast to~\eqref{firstlaw0}, we find that the contribution from scalar hair has a particularly simple and instructive form, from which we are able to give a clear physical interpretation for both $X$ and $Y$. We also construct the hairy black holes explicitly to demonstrate their existence and to numerically check the resulting thermodynamic relations. Finally, a general argument is given for the form of scalar hair contribution to the first law of thermodynamics.

\section{Setup and Background}
We are interested in the thermodynamics of the (d+1)-dimensional asymptotically AdS black holes in Einstein-scalar theory~\eqref{action}. We consider the planar black holes with the ansatz given by
\begin{equation}\label{phiansatz}
ds^2=-f(r) e^{-\eta(r)} dt^2+\frac{dr^2}{f(r)}+r^2 d\vec{x}_{d-1}^2,\quad \phi=\phi(r)\,,
\end{equation}
for which the location of the event horizon is at $r=r_h$ where $f(r_h)=0$.  We take the geometry at the black hole boundary $r\rightarrow \infty$ to be $AdS_{d+1}$. Although we focus on the case with planar symmetry in the present study, our discussion below should apply to black holes with spherical and hyperbolic horizon geometries.

Near the AdS boundary, one has $f(r\rightarrow \infty)=r^2/L^2$, with $L$ the AdS curvature scale, and $\eta(r\rightarrow \infty)=0$.\,\footnote{In general one only needs $\eta$ to approach a constant near the AdS boundary. But it should not be thought of as a free parameter, as it can be absorbed into a rescaling of the time coordinate. We choose $\eta(r\rightarrow \infty)=0$ such that the asymptotic AdS metric has a
canonically normalized time coordinate, \emph{i.e.} we are interested in the boundary system with respect to the metric $ds^2=-dt^2+d\vec{x}_{d-1}^2$.} Without loss of generality, near the AdS boundary we parameterize the scalar potential as follows
\begin{equation}
V(\phi)=-\frac{d(d-1)}{L^2}-\frac{1}{2}m^2\phi^2+\mathcal{O}(\phi^3)\,,
\end{equation}
where we have chosen $\phi\rightarrow 0$ at the black hole boundary for convenience, and the parameter $m^2$ is the mass of the scalar field. Then the asymptotic expansion for the scalar near the AdS boundary is schematically given by
\begin{equation}\label{UVphi}
\phi(r)=\frac{\phi_s}{r^{d-\Delta}}(1+\dots)+\frac{\phi_v}{r^{\Delta}}(1+\dots)\,,
\end{equation}
where $\Delta=(d+\sqrt{d^2+4m^2 L^2})/2$, and $\phi_s$ and $\phi_v$ are two constants. In order to allow the most general solutions with two non-trivial parameters $\phi_s$ and $\phi_v$, we have considered the case where the mass-squared of the scalar field is negative, but above the Breitenlohner-Freedman (BF) bound $m_{BF}^2=-\frac{d^2}{4 L^2}$,  \emph{i.e.} $m_{BF}^2<m^2<0$.\,\footnote{There is a special case saturating the BF bound with $m^2=-\frac{d^2}{4 L^2}$, for which the scalar asymptotics behaves different from~\eqref{UVphi}. This case does not change our discussion below and a concrete example is presented in the appendix.} Following the spirit of holography, we call the leading term $\phi_s$ as the source.\,\footnote{For the mass range slightly above the BF bound, $-\frac{d^2}{4}<m^2L^2<1-\frac{d^2}{4}$, the two falloffs of~\eqref{UVphi} are both normalizable and either one can be considered as a source, corresponding to two different dual field theories~\cite{Klebanov:1999tb}. We will only consider standard quantization for the scalar field throughout the manuscript.}
In the rest of our discussion, without loss of generality, we shall work with $L = 1$ for simplicity by fixing in such a way the length unit. 

The equations of motion that follow from the action~\eqref{action} are:
\begin{equation}\label{eomphi}
\nabla_\mu \nabla^\mu\phi-\partial_\phi V=0,
\end{equation}
\begin{equation}
\mathcal{R}_{\mu\nu} -\frac{1}{2}\mathcal{R}g_{\mu\nu}= \frac{1}{2}\partial_\mu\phi \partial_\nu\phi+\frac{1}{2}\left(-\frac{1}{2}\nabla_\mu \phi \nabla^\mu \phi-V\right)g_{\mu\nu}\,.
\end{equation}
Substituting the ansatz~\eqref{phiansatz} into the above equations of motion, we obtain the following equations of motion:
\begin{equation}\label{Eoms}
\begin{split}
\phi''+\left(\frac{f'}{f}-\frac{\eta'}{2}+\frac{d-1}{r}\right)\phi'-\frac{1}{f}\partial_\phi V = &0\,, \\
\frac{\eta'}{r}+\frac{1}{d-1}\phi'^2=&0\,,\\
\frac{2}{r}\frac{ f'}{f}-\frac{\eta'}{r}+\frac{2}{d-1}\frac{V}{f}+\frac{2(d-2)}{r^2}=&0\,,\\
\frac{f''}{f}-\eta''+\frac{1}{2}\eta'^2+\phi'^2+\left(\frac{d-3}{r}-\frac{3}{2}\eta'\right)\frac{f'}{f}-\frac{2(d-2)}{r^2}=&0\,,
\end{split}
\end{equation}
where only two of the last three equations are independent.
The temperature and entropy density associated with the above background are given by
\begin{equation}\label{tems}
T=\frac{1}{4\pi}f'(r_h)e^{-\eta(r_h)/2},\qquad s=\frac{2\pi}{\kappa_N^2} r_h^{d-1}\,,
\end{equation}
respectively. 

Making use of~\eqref{Eoms}, we can obtain a radial conserved quantity\,\footnote{For black holes with spherical and hyperbolic horizon topologies, there is an extra term in $\mathcal{Q}$  due to non-vanishing curvature of horizons~\cite{Cai:2020wrp}.}
\begin{equation}\label{eomQ}
\mathcal{Q}=r^{d+1} e^{\eta/2}\left(\frac{f}{r^2}e^{-\eta}\right)'\,,
\end{equation}
which connects horizon to boundary data. Evaluating~\eqref{eomQ} at the horizon $r=r_h$ and using~\eqref{tems}, one finds
\begin{equation}
\mathcal{Q}=2\kappa_N^2\, T s\,.
\end{equation}
Therefore, $\mathcal{Q}=0$ signals extremity. As a consequence, for the Einstein-scalar theory the extremality condition implies $\frac{f(r)}{r^2}e^{-\eta(r)}$ is a constant, and thus extremal geometries are relativistic independently of which potential $V(\phi)$ one chooses~\cite{Cremonini:2020rdx}. The extremal geometry is described by a metric of the form
\begin{equation}
ds^2=\frac{dr^2}{f(r)}+r^2 (-dt^2+d\vec{x}_{d-1}^2)\,,
\end{equation}
where for convenience we have set $\frac{f(r)}{r^2}e^{-\eta(r)}=1$. In order to compute $\mathcal{Q}$ at the AdS boundary $r\rightarrow\infty$, one needs to know the boundary asymptotics for the bulk fields $f$ and $\eta$, which depend on the details of the theory one considers.

\section{Wald Formula}\label{sec:wald}

Our next step is to consider the first law of thermodynamics for our Einstein-scalar black holes~\eqref{phiansatz}. We will adopt the general procedure developed by Wald~\cite{Wald:1993nt}. In Wald's procedure, one considers the variation of the parameters in a (d+1) dimensional solution, and constructs a closed (d-1)-form $\delta Q-i_\xi\Theta$ with $\xi$ a Killing vector. To make contact with the first law of black hole thermodynamics, we take $\xi=\partial/\partial_t$, \emph{i.e.} the time-like Killing vector that is null on the horizon. Then one introduces the integral
\begin{equation}
\delta H=\int_{\Sigma^{(d-1)}}(\delta Q-i_\xi\Theta)\,,
\end{equation}
over any (d-1)-dimensional surface $\Sigma^{(d-1)}$ at constant $t$ and $r$.  It has been shown by~\cite{Wald:1993nt} that the variation of the Hamiltonian $\delta H$ is independent of $r$. 

For our present case~\eqref{phiansatz} of the Einstein-scalar theory, we follow the derivation presented in~\cite{Liu:2013gja,Lu:2014maa} and obtain that
\begin{equation}
\begin{split}
Q&=\frac{1}{2\kappa_N^2} r^{d-1}(fe^{-\eta})'e^{\eta/2}\,dx_1\wedge dx_2\wedge\dots\wedge dx_{d-1}\,,\\
i_{\xi}\Theta&=\frac{1}{2\kappa_N^2} r^{d-1}\left[\delta((fe^{-\eta})'e^{\eta/2})+\frac{d-1}{r}e^{-\eta/2}\delta f+fe^{-\eta/2}\phi'\delta\phi\right]\,dx_1\wedge dx_2\wedge\dots\wedge dx_{d-1}\,.
\end{split}
\end{equation}
Then $\delta H$ at radius $r$ is given by
\begin{equation}\label{eomH}
\delta H=-\frac{\Sigma}{2\kappa_N^2} r^{d-1} e^{-\eta/2}\left[\frac{d-1}{r}\delta f+f\phi'\delta\phi\right]\,,
\end{equation}
with $\Sigma$ the spatial volume of the (d-1)-plane. Since $\delta H$ is a radially conserved quantity, it can be evaluated at any position. Near the horizon at $r=r_h$, we have
\begin{equation}
f(r)=f'(r_h)(r-r_h)+\mathcal{O}(r-r_h)^2,\quad \eta(r)=\eta(r_h)+\mathcal{O}(r-r_h),\quad \phi(r)=\phi(r_h)+\mathcal{O}(r-r_h)\,.
\end{equation}
Thus $\delta f|_{r=r_h}=-\delta r_h f'(r_h)$, and we obtain
\begin{equation}\label{horizonH}
\frac{\delta H_h}{\Sigma}=\frac{1}{2\kappa_N^2} e^{-\eta(r_h)/2}f'(r_h)(d-1) r_h^{d-2}\delta r_h= T\delta s\,,
\end{equation}
where we have used $\delta s=\delta(\frac{2\pi}{\kappa_N^2}r_h^{d-1})=\frac{2\pi}{\kappa_N^2}(d-1)r_h^{d-2}\delta r_h$. 

In order to evaluate~\eqref{eomH} at the boundary, one should know the boundary asymptotics for three bulk fields $f$, $\eta$ and $\phi$. The expansions near the black hole boundary depend on the precise form of $V(\phi)$. For a given potential $V(\phi)$, one can in principle obtain the full asymptotic form by taking appropriate large-$r$ expansions for $f$, $\eta$ and $\phi$, inserting them into the equations of motion~\eqref{Eoms}, and solving for the coefficients in the expansions up to some desired order. In practice, it is in general difficult to make the appropriate expansions due to the backreaction of the scalar to the metric. In certain cases, there is logarithmic $r$ dependence in the asymptotic expansions for the metric and scalar fields. 

Furthermore, we also need to obtain the mass of the Einstein-scalar black holes, for which we use the standard holographic techniques. More precisely, we calculate the renormalized stress tensor $T_{\mu\nu}$ for the dual boundary theory and the mass is obtained from the $tt$ component of the holographic stress tensor, \emph{i.e.} $\mathcal{E}=T_{tt}$. In order to do that, we should add appropriate boundary terms and counterterms to the bulk action to remove divergences in the holographic stress tensor, which also depends on the non-linear detail of the coupling $V(\phi)$. To avoid above complications, our strategy is to consider some  representative ``benchmark" models in this work. We will find the precise form of boundary asymptotics and will examine the thermodynamics of the black hole solutions via the holographic renormalization. 

\section{Five Dimensional Case}
For the five-dimensional case, we consider the following class of potentials:
\begin{equation}\label{vpotential}
V(\phi)=(6\gamma^2-\frac{3}{2})\phi^2-12\cosh[\gamma\phi]\,.
\end{equation}
with $\gamma$ a constant. Near the AdS boundary where $\phi\rightarrow 0$, one has
\begin{equation}
V(\phi)=-12-\frac{3}{2}\phi^2+\mathcal{O}(\phi^4)\,.
\end{equation}
Therefore, the the cosmological constant is given by $\Lambda=-6/L^2$ with the AdS radius $L=1$ and $\Delta=3$ of~\eqref{UVphi}.  

Using the equations of motion~\eqref{Eoms}, the boundary asymptotics as $r\rightarrow\infty$ for the bulk fields $f(r), \eta(r)$ and $\phi(r)$ are found to be
\begin{equation}\label{UVexpand5D}
\begin{split}
\phi(r)& = \frac{\phi_s}{r}+\frac{\phi_v}{r^3}-\frac{\ln(r)}{6r^3}(1-6\gamma^4)\phi_s^3+\mathcal{O}(\frac{\ln(r)}{r^5})\,, \\
f(r)&=r^2\left[1+\frac{\phi_s^2}{6r^2}+\frac{f_v}{r^4}-\frac{\ln(r)}{12 r^4}(1-6\gamma^4)\phi_s^4+\mathcal{O}(\frac{\ln(r)^2}{r^6})\right]\,,\\
\eta(r)&=\eta_0+\frac{\phi_s^2}{6r^2}+\frac{(1-6\gamma^4)\phi_s^4+72\phi_s\phi_v}{144 r^4}-\frac{\ln(r)}{12 r^4}(1-6\gamma^4)\phi_s^4+\mathcal{O}(\frac{\ln(r)^2}{r^6})\,.
\end{split}
\end{equation}
We point out that due to the non-trivial source $\phi_s$, there are terms with $\ln(r)$ dependence in the above expansion.
As we have pointed out, we will set $\eta_0=0$ to fix the normalization of time.

\subsection{Thermodynamic relation}

To read off the physical observables via the holographic renormalization, we need to consider appropriate boundary terms in the UV. 
For the present class of models with $V$ given in~\eqref{vpotential}, the boundary terms are found to be
\begin{eqnarray}\label{ct5D}
S_\partial=\frac{1}{2\kappa_N^2}\int_{r\rightarrow\infty}dx^4\sqrt{-h}\left[2K-6-\frac{1}{2}\phi^2-\frac{6\gamma^4-1}{12}\phi^4 \ln(r)-b\,\phi^4\right]\,,
\label{expansion}\end{eqnarray}
where $h_{\mu\nu}$ is the induced metric at the AdS boundary and $K_{\mu\nu}$ the extrinsic curvature defined by the outward pointing normal vector to the boundary between $n^\mu$. The first term of~\eqref{ct5D} is Gibbons-Hawking term for a well-defined Dirichlet variational principle, and the three local covariant surface counterterms in the middle are required to remove divergences. The last term is a finite counterterm with $b$ typically a constant free parameter, which defines a renormalization scheme. As will be demonstrated below, most of the thermodynamic variables depend on the parameter $b$.  Note also that the coefficient of the logarithmic term depends on the model parameter $\gamma$.
According to the holographic dictionary, the holographic stress tensor is
\begin{eqnarray}\label{Tmunu}
T_{\mu\nu}=\frac{1}{2\kappa_N^2}\lim_{r\rightarrow\infty}r^2\left[2(K h_{\mu\nu}-K_{\mu\nu}-3 h_{\mu\nu})-\left(\frac{1}{2}\phi^2+\frac{6\gamma^4-1}{12}\phi^4 \ln(r)+b\,\phi^4\right)h_{\mu\nu}\right]\,,
\label{expansion}\end{eqnarray}
and the condensate for the dual scalar operator reads
\begin{equation}\label{Vev}
\left<O\right>=\frac{\delta}{\delta\phi_s}(S+S_\partial)_{on-shell}=\frac{1}{2\kappa_N^2}\lim_{r\rightarrow\infty}\frac{\sqrt{-h}}{r}\left[-n_\mu \nabla^\mu \phi-\phi-\frac{6\gamma^4-1}{3}\phi^3\ln(r)-4b\,\phi^3\right]\,.
\end{equation}

Inserting the boundary expansions of the bulk fields~\eqref{UVexpand5D} into~\eqref{Tmunu}, we obtain the non-vanishing components of the stress tensor
\begin{equation}\label{EP}
\begin{split}
2\kappa_N^2 T_{tt}=2\kappa_N^2\mathcal{E}&=-3 f_v+\phi_s\phi_v+\frac{1+48 b}{48}\phi_s^4\,,\\
2\kappa_N^2 (T_{xx}=T_{yy}=T_{zz})=2\kappa_N^2 P&=-f_v+\phi_s\phi_v+\frac{3-48 b-8\gamma^4}{48}\phi_s^4\,,
\end{split}
\end{equation}
where $P$ is the pressure and  $\mathcal{E}$ is the energy density which is also known as the mass density of the black hole.\,\footnote{In our discussion the boundary geometry possesses a time-like Killing vector $\xi^\mu=(\frac{\partial}{\partial t})^\mu$. We define the energy density of the dual field theory as $\mathcal{E}=T_{\mu\nu}\xi^\mu\xi^\nu$, which is equal to the energy density of the black hole via the AdS/CFT correspondence. Instead of energy $E$ and entropy $S$ of~\eqref{firstlaw0}, we use energy density $\mathcal{E}$ and entropy density $s$ in the planar black hole case~\eqref{phiansatz}.}
 The trace of the stress tensor is given by
\begin{equation}
{T^\mu}_\mu=2\kappa_N^2(3P-\mathcal{E})=2\phi_s\phi_v+\frac{1-24 b-3\gamma^4}{6}\phi_s^4\,.
\end{equation}
Note that there is a trace anomaly in the presence of the source term $\phi_s$ which breaks the conformal symmetry. The expectation value of the dual scalar operator obtained from~\eqref{Vev} reads
\begin{equation}\label{condensate}
\left<O\right>=\frac{1}{2\kappa_N^2}\left(2\phi_v+\frac{1-16 b-4\gamma^4}{4}\phi_s^3\right)\,.
\end{equation}
It is clear that $\mathcal{E}$, $P$ and $\left<O\right>$ depend on the parameter $b$ of~\eqref{ct5D}, corresponding to the renormalization scheme.

We now compute the free energy density $\Omega$ which is identified as the temperature $T$ times the renormalized action in Euclidean signature. Since we consider a stationary problem, the Euclidean action is related to the Minkowski one by a minus sign. Therefore, we have
\begin{equation}\label{freeE}
-\Omega \Sigma=T (S+S_\partial)_{on-shell}\,,
\end{equation}
with $\Sigma$ the spatial volume and $t\in[0,1/T]$. Employing the equations of motion, we find
\begin{equation}
\Omega=\lim_{r\rightarrow\infty}\frac{1}{2\kappa_N^2}\left[2 e^{-\eta/2}r^2f-e^{-\eta/2}r^3\sqrt{f}\left(2K-6-\frac{1}{2}\phi^2-\frac{6\gamma^4-1}{12}\phi^4\ln(r)-b\phi^4\right)\right]\,,
\end{equation}
Substituting the asymptotical expansion~\eqref{UVexpand5D} into above expression, we obtain the free energy
\begin{equation}
\Omega=\frac{1}{2\kappa_N^2}\left(f_v-\phi_s\phi_v-\frac{3-48 b-8\gamma^4}{48}\phi_s^4\right)\,.
\end{equation}
One immediately finds that the pressure reappears in the analysis as just minus the free energy, \emph{i.e.} $P=-\Omega$, which is expected from thermodynamics.
Furthermore, evaluating the conserved quantity $\mathcal{Q}$~\eqref{eomQ} at the AdS boundary, one obtains
\begin{equation}
\mathcal{Q}=-4 f_v+2\phi_s\phi_v+\frac{1-2\gamma^4}{12}\phi_s^4=2\kappa_N^2(\mathcal{E}+P)\,,
\end{equation}
where we have used~\eqref{EP}. So we manage to obtain the expected thermodynamic relation
\begin{equation}\label{Free5D}
\Omega=\mathcal{E}-T\,s=-P\,,
\end{equation}
for the class of potentials in~\eqref{vpotential}, independent of the renormalization scheme one chooses.

\subsection{First law of thermodynamics}
We now consider the first law of thermodynamics for our Einstein-scalar black hole by considering the Wald formula. In the present case with $d=4$, we compute the radial conserved quantity
\begin{equation}\label{eomH5D}
\delta H=-\frac{\Sigma}{2\kappa_N^2} r^3 e^{-\eta/2}\left[\frac{3}{r}\delta f+f\phi'\delta\phi\right]\,,
\end{equation}
both at the horizon and the AdS boundary.

As we have already shown in~\eqref{horizonH}, the horizon computation yields
\begin{equation}
\frac{\delta H_h}{\Sigma}=\frac{1}{2\kappa_N^2} e^{-\eta(r_h)/2}f'(r_h)3 r_h^2\delta r_h= T\delta s\,.
\end{equation}
Substituting the UV expansion~\eqref{UVexpand5D} into~\eqref{eomH5D} and evaluating it at infinity, we obtain
\begin{equation}
\begin{split}
\frac{\delta H_\infty}{\Sigma}&=\frac{1}{2\kappa_N^2}\left[-3\delta f_v+\phi_s\delta\phi_v+\frac{(1-3\gamma^4)\phi_s^3+3\phi_v}{3}\delta\phi_s\right]\\
&=\delta\mathcal{E}+\frac{1}{2\kappa_N^2}\left(2\phi_v+\frac{1-16 b-4\gamma^4}{4}\phi_s^3\right)\delta\phi_s=\delta\mathcal{E}+\left<O\right>\delta\phi_s\,,
\end{split}
\end{equation}
where the expressions for energy density~\eqref{EP} and condensate~\eqref{condensate} have been used in the second and third equalities, respectively. Therefore, we have the following first law of thermodynamics
\begin{equation}\label{firstlaw}
d\mathcal{E}=T d s-\left<O\right>d\phi_s\,,
\end{equation}
and in terms of free energy~\eqref{Free5D} we have
\begin{equation}
d\Omega=-s d T-\left<O\right>d\phi_s\,.
\end{equation}
Although the boundary behaviors of bulk fields are sensitive to the model parameter $\gamma$ and the thermodynamic quantities, such as $\mathcal{E}$ and $\left<O\right>$, depend on the choice of parameter $b$ in the boundary counterterms, the first law of thermodynamics can be written as above simple and compact form.

\begin{figure}
\begin{center}
\includegraphics[width=2.8in]{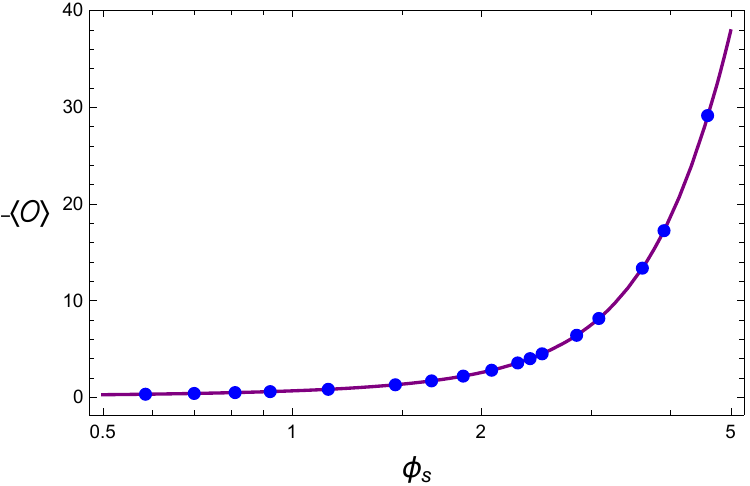}\quad
\includegraphics[width=2.73in]{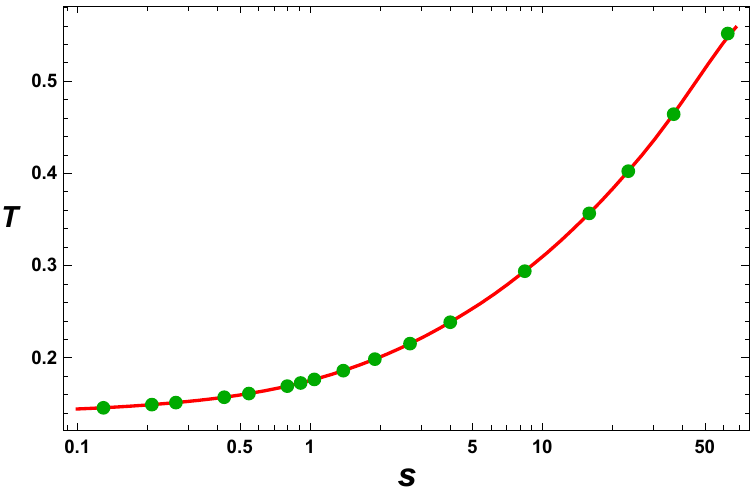}
\caption{Test the first law of thermodynamics $d\mathcal{E}=T d s-\left<O\right>d\phi_s$ for the five dimensional hairy black hole solutions with the potential~\eqref{vpotential}. \textbf{Left:} Fix the entropy density $s=4\pi$. The solid purple curve is obtained via the first law $\left<O\right>=-\left(\frac{\partial \mathcal{E}}{\partial \phi_s}\right)_s$, while the blue dots are read off directly from the AdS boundary~\eqref{condensate}. The behavior of $\left<O\right>$ depends on the parameter $b$ for which we take $b=0.05$ as an example. \textbf{Right:} Fix the source $\phi_s=1$. The solid red curve is obtained by the first law $T=\left(\frac{\partial \mathcal{E}}{\partial s}\right)_{\phi_s}$, and the green dots are from the horizon data by using~\eqref{tems}. The agreement is excellent. We have fixed $2\kappa_N^2=1$ and chosen $\gamma=0.7$.}
\label{fig:5D}
\end{center}
\end{figure}

To check the first law of thermodynamics, we construct the hairy black holes by solving the equations of motion~\eqref{Eoms} numerically for the potentials~\eqref{vpotential} with $d=4$. We can then read off $(\phi_s, \left<O\right>, \mathcal{E}, P)$ from the black hole boundary~\eqref{UVexpand5D}, and obtain $(T, s)$ from the horizon data via~\eqref{tems}. The presence of logarithmic terms in~\eqref{UVexpand5D} poses some technical complications on extracting physical data. In practice, we fit the numerical solutions close to the AdS boundary using ultraviolet asymptotics~\eqref{UVexpand5D} and obtain $(\phi_s,\phi_v, f_v)$ which are required to calculate most of the physical quantities. According to the first law~\eqref{firstlaw}, one should have $\left<O\right>=-\left(\frac{\partial \mathcal{E}}{\partial \phi_s}\right)_s$ with the entropy density $s$ fixed and $T=\left(\frac{\partial \mathcal{E}}{\partial s}\right)_{\phi_s}$ with the source $\phi_s$ fixed. In figure~\ref{fig:5D}, we compare $\left<O\right>$ and $T$ extracted from the boundary data and the first law of thermodynamics~\eqref{firstlaw}. The results from both ways match with each other perfectly. We also check the thermodynamic relation~\eqref{Free5D}, which is found to be satisfied with high numerical accuracy.

\section{Four Dimensional Case}
In this section we consider a family of models in four dimensions. The potential reads
\begin{equation}\label{vpotential4D}
V(\phi)=-6-\frac{4}{\delta^2}\sinh\left[\frac{\delta\phi}{2}\right]^2\,,
\end{equation}
with $\delta$ a free parameter.\footnote{This particular form of scalar potential was introduced in~\cite{Kiritsis:2015oxa}. It allows the near boundary expansion that is consistent without the logarithmic terms.} Near the AdS boundary where $\phi\rightarrow 0$, one has
\begin{equation}
V(\phi)=-6-\phi^2+\mathcal{O}(\phi^4)\,,
\end{equation}
from which one obtains the cosmological constant $\Lambda=-3$ and the parameter $\Delta=2$ of~\eqref{UVphi}. The asymptotics for the three bulk fields $f(r), \eta(r)$ and $\phi(r)$ near the AdS boundary are given by
\begin{equation}\label{UVexpand4D}
\begin{split}
\phi(r)& = \frac{\phi_s}{r}+\frac{\phi_v}{r^2}-\frac{(3-4\delta^2)\phi_s^3}{24\, r^3}+\mathcal{O}(\frac{1}{r^4})\,, \\
f(r)&=r^2\left[1+\frac{\phi_s^2}{4\,r^2}+\frac{f_v}{r^3}+\mathcal{O}(\frac{1}{r^4})\right]\,,\\
\eta(r)&=\frac{\phi_s^2}{4\,r^2}+\frac{2\phi_s\phi_v}{3\, r^3}+\mathcal{O}(\frac{1}{r^4})\,,
\end{split}
\end{equation}
where we have set $\eta(r\rightarrow\infty)=0$ to fix the normalization of time. In contrast to the five dimensional case, there are no logarithmic terms in above expansions due to the particular form of potential in~\eqref{vpotential4D}.

\subsection{Thermodynamic relation}
Following the holographic renormalization procedure, we introduce the counterterm action for the class of potentials~\eqref{vpotential4D}
\begin{eqnarray}\label{ct4D}
S_\partial=\frac{1}{2\kappa_N^2}\int_{r\rightarrow\infty}dx^3\sqrt{-h}\left[2K-4-\frac{1}{2}\phi^2-c\,\phi^3\right]\,,
\label{expansion}\end{eqnarray}
where the first one is Gibbons-Hawking boundary term for a well-defined Dirichlet variational principle and the following two terms for removing divergence. The last term of~\eqref{ct4D} is a finite counterterm with $c$ a constant, corresponding to the freedom to choose the counterterms in holographic renormalization. 
Then we obtain the holographic stress tensor
\begin{eqnarray}\label{Tmunu4D}
T_{\mu\nu}=\frac{1}{2\kappa_N^2}\lim_{r\rightarrow\infty}r\left[2(K h_{\mu\nu}-K_{\mu\nu}-2 h_{\mu\nu})-\left(\frac{1}{2}\phi^2+c\,\phi^3\right)h_{\mu\nu}\right]\,,
\label{expansion}\end{eqnarray}
and the condensate for the dual scalar operator reads
\begin{equation}\label{Vev4D}
\left<O\right>=\frac{\delta}{\delta\phi_s}(S+S_\partial)_{on-shell}=\frac{1}{2\kappa_N^2}\lim_{r\rightarrow\infty}\frac{\sqrt{-h}}{r}\left[-n_\mu \nabla^\mu \phi-\phi-3c\,\phi^2\right]\,.
\end{equation}

By substituting the expansions~\eqref{UVexpand4D} into~\eqref{Tmunu4D}, we obtain the stress tensor
\begin{equation}\label{EP4D}
\begin{split}
2\kappa_N^2 T_{tt}=2\kappa_N^2\mathcal{E}&=-2 f_v+\phi_s\phi_v+c\phi_s^3\,,\\
2\kappa_N^2 (T_{xx}=T_{yy}=T_{zz})=2\kappa_N^2 P&=-f_v+\phi_s\phi_v-c \phi_s^3\,,
\end{split}
\end{equation}
where $P$ is the pressure and $\mathcal{E}$ is the energy (mass) density. Due to the presence of the source term $\phi_s$, the stress tensor also has a non-vanishing trace
\begin{equation}
2\kappa_N^2(2P-\mathcal{E})=\phi_s\phi_v-3 c \phi_s^3\,.
\end{equation}
The expectation value of the dual scalar operator obtained from~\eqref{Vev4D} reads
\begin{equation}\label{condensate4D}
\left<O\right>=\frac{1}{2\kappa_N^2}\left(\phi_v-3 c \phi_s^2\right)\,.
\end{equation}

Similar to the five dimensional case, we can obtain the free energy
\begin{equation}
\Omega=\frac{1}{2\kappa_N^2}(f_v-\phi_s\phi_v+c \phi_s^3)=-P\,.
\end{equation}
Evaluating the conserved quantity $\mathcal{Q}$~\eqref{eomQ} at the AdS boundary, one obtains
\begin{equation}
\mathcal{Q}=-3 f_v+2\phi_s\phi_v=2\kappa_N^2(\mathcal{E}+P)\,,
\end{equation}
where we have used~\eqref{EP4D}. Therefore, we manage to obtain the expected thermodynamic relation
\begin{equation}\label{Free4D}
\Omega=\mathcal{E}-T\,s=-P\,.
\end{equation}
for the four dimensional hairy black holes.

\subsection{First law of thermodynamics}
We now consider the first law of thermodynamics by using the Wald formula developed in Section~\ref{sec:wald}. For the present case with $d=3$, we compute the radial conserved quantity $\delta H$~\eqref{eomH} at the AdS boundary
and obtain that
\begin{equation}
\frac{\delta H_\infty}{\Sigma}=-\frac{1}{2\kappa_N^2}\lim_{r\rightarrow\infty} r^{2} e^{-\eta/2}\left[\frac{2}{r}\delta f+f\phi'\delta\phi\right]=\frac{1}{2\kappa_N^2}[-2\delta f_v+\phi_s\delta\phi_v+2\phi_v\delta\phi_s]\,.
\end{equation}
Note that the variation of the energy density is given from~\eqref{EP4D} that
\begin{equation}
\delta\mathcal{E}=\frac{1}{2\kappa_N^2}[-2\delta f_v+\phi_s\delta\phi_v+(\phi_v+3c\phi_s^2)\delta\phi_s]\,.
\end{equation}
Then we find 
\begin{equation}
\frac{\delta H_\infty}{\Sigma}=\delta\mathcal{E}+\frac{1}{2\kappa_N^2}\left(\phi_v-3 c \phi_s^3\right)\delta\phi_s=\delta\mathcal{E}+\left<O\right>\delta\phi_s\,,
\end{equation}
where the last equation is obtained by using the expression for the condensate~\eqref{condensate4D}. Therefore, we have the following first law of thermodynamics
\begin{equation}\label{firstlaw4D}
d\mathcal{E}=T d s-\left<O\right>d\phi_s\,,
\end{equation}
and in terms of free energy we have
\begin{equation}
d\Omega=-s d T-\left<O\right>d\phi_s\,.
\end{equation}
Although $\mathcal{E}$ and $\left<O\right>$ depend on the parameter $c$ in the boundary counterterms~\eqref{vpotential4D}, the final form of the first law of thermodynamics is independent of $c$.

\begin{figure}
\begin{center}
\includegraphics[width=2.8in]{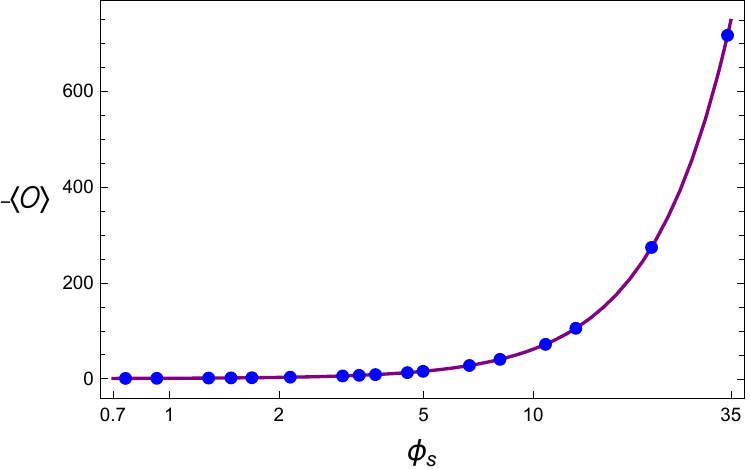}\quad
\includegraphics[width=2.73in]{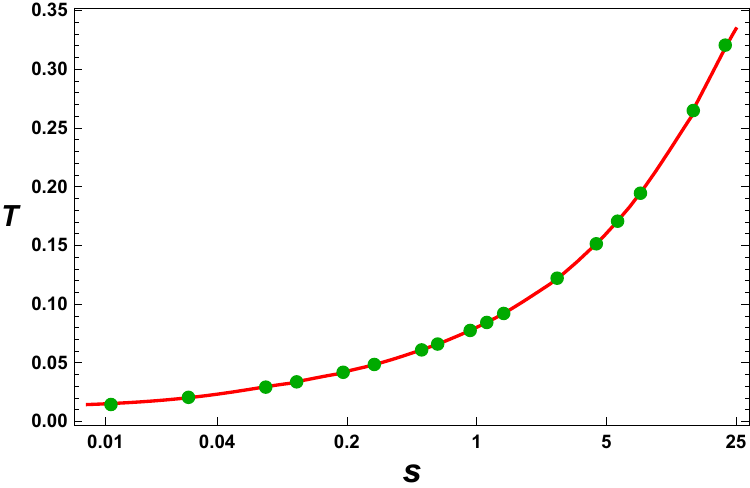}
\caption{Check the first law of thermodynamics $d\mathcal{E}=T d s-\left<O\right>d\phi_s$ for the four dimensional hairy black hole solutions with the potential~\eqref{vpotential4D}. \textbf{Left:} Fix the entropy density $s=4\pi$. The solid purple curve is obtained via the first law $\left<O\right>=-\left(\frac{\partial \mathcal{E}}{\partial \phi_s}\right)_s$, while the blue dots are extracted from the boundary data~\eqref{condensate4D}. The behavior of $\left<O\right>$ depends on the parameter $c$ for which we take $c=0.1$ as an example. \textbf{Right:} Fix the source $\phi_s=1$. The solid red curve is obtained by the first law $T=\left(\frac{\partial \mathcal{E}}{\partial s}\right)_{\phi_s}$, and the green dots are read off from the horizon data by using~\eqref{tems}. The agreement is manifest. We have fixed $2\kappa_N^2=1$ and chosen $\delta=0.5$.}
\label{fig:4D}
\end{center}
\end{figure}

In order to further check our discussion, we construct the four dimensional hairy black holes numerically for the potentials~\eqref{vpotential4D}, and read off thermodynamic quantities $(\phi_s, \phi_v,\left<O\right>, \mathcal{E}, T, s)$ from boundary data via~\eqref{tems},~\eqref{EP4D} and~\eqref{condensate4D}. The thermodynamic relation~\eqref{Free4D} is found to be satisfied with very high numerical accuracy. In figure~\ref{fig:4D} we show the observables extracted  from the first law of thermodynamics~\eqref{firstlaw4D}, including $\left<O\right>=-\left(\frac{\partial \mathcal{E}}{\partial \phi_s}\right)_s$ (left panel) and $T=\left(\frac{\partial \mathcal{E}}{\partial s}\right)_{\phi_s}$ (right panel). Both exactly agree with the ones obtained from boundary data directly.

\section{Discussion}
We have examined the thermodynamics for black holes with non-trivial scalar hair in Einstein gravity coupled to a scalar field in asymptotically AdS spacetime. We defined the energy or mass of the hairy black hole from the holographic stress tensor via the standard holographic renormalization. Although the explicit expressions for thermodynamic variables, such as $\mathcal{E}$ and $P$, depend on the details of a model one considers, we have shown that the thermodynamic relation and the first law of thermodynamics can be written as the model-independent language already advocated in terms of thermodynamic quantities, see \emph{e.g.}~\eqref{Free5D} and~\eqref{firstlaw}.
More precisely, by incorporating  the Wald formalism, we have found that the first law of thermodynamics is modified with a novel contribution from the scalar hair. One recovers the traditional first law of thermodynamics $d\mathcal{E}=T ds$ by restricting on the family of solutions with the source $\phi_s$ fixed, \emph{i.e.} $d\phi_s=0$. We have checked numerically the new form of first law and thermodynamic relations for two families of models in five and four dimensions.\,\footnote{Although a scalar is the simplest field in any field theory, it is still difficult to obtain analytic scalar black hole solutions with general coefficients $\phi_s$ and $\phi_v$ of~\eqref{UVphi}. There have been some analytic hairy black holes, see \emph{e.g.}~\cite{Anabalon:2012ta,Gonzalez:2013aca,Anabalon:2013sra,Feng:2013tza,Ren:2019lgw},  but these are very special solutions for which the first law of thermodynamics does not contain any contribution from the scalar hair.} The special case that saturates the BF bound was also examined in the appendix. 

The new form of the first law of thermodynamics~\eqref{firstlaw} seems quite natural from the holographic point of view. Identifying the leading term $\phi_s$ of the scalar $\phi$ in the boundary expansion~\eqref{UVphi} as the source, we introduce its conjugate $\left<O\right>$ using the GKPW formula~\cite{Gubser:1998bc,Witten:1998qj}
\begin{equation}
\left<O\right>=\frac{\delta}{\delta\phi_s}(S+S_\partial)_{on-shell}\,.
\end{equation}
On the other hand, for a stationary background geometry, the renormalized on-shell action $(S+S_\partial)_{on-shell}$ is closely related to the free energy via~\eqref{freeE}. Therefore, we immediately obtain
\begin{equation}
\left<O\right>\equiv\frac{\delta}{\delta\phi_s}(S+S_\partial)_{on-shell}=-\frac{\partial \Omega}{\partial\phi_s}=-\frac{\partial \mathcal{E}}{\partial\phi_s}\,,
\end{equation}
where we have used $\mathcal{E}=\Omega-T s$ in the last equality. Thus, it is reasonable to consider $\phi_s$ and $\left<O\right>$ as a pair of thermodynamic conjugate variables in the first law.

It has been known that in the presence of a scalar hair there should be a non-trivial contribution to the first law of thermodynamics~\eqref{Zterm}. While this result has been checked in many examples~\cite{Liu:2013gja,Lu:2014maa,Lu:2013ura}, a deeper and broader physical understanding is needed for this new correction. Using the standard holographic dictionary, our analysis suggests that such contribution has a particular form $\delta\mathcal{E}=-\left<O\right>\delta \phi_s$ where $\left<O\right>$ is the conjugate of $\phi_s$ in terms of the renormalized on-shell action and is precisely the response of the dual operator in the presence of the source $\phi_s$. Comparing our result $d\mathcal{E}=T d s-\left<O\right>d\phi_s$ with~\eqref{firstlaw0}, we can give a clear physical interpretation on the pair of thermodynamic variables $(X, Y)$: $Y$ is the source term of the scalar hair and $X$ the response of the scalar from the holographic point of view. In particular, for the branch of solutions with the source $Y\equiv\phi_s$ fixed, which is the typical situation for holographic applications, such additional contribution to the first law vanishes. It also means that only for the branch of black hole solutions with $Y\equiv\phi_s$ fixed can one obtain the pressure by directly integrating the first law of thermodynamics, \emph{i.e.} $P=\int s\, dT$.

We point out that there is some subtlety on the definition of mass for asymptotically AdS black holes. In the present work we have defined the mass using standard holographic techniques. Apart from the holographic mass, other definitions include ADM mass~\cite{Ashtekar:1984zz,Ashtekar:1999jx}, Komar mass~\cite{Barnich:2004uw,Kastor:2009wy}, Hamiltonian mass~\cite{Lu:2013ura,Chow:2013gba} and so on. A thorough study on the role of different definitions of mass in the first law of thermodynamics is interesting but beyond the scope of this manuscript.
Although we did not consider all possible models, our discussion suffices to illustrate the key feature we uncover. It should be helpful to examine other cases of potential with a different scalar mass.
In the present study we focused on the neutral case, but our discussion can be easily extended to the charged case by adding a U(1) gauge field. It would also apply to other kinds of matter content, such as massive vector and p-from fields, by incorporating the Wald formalism and the holographic renormalization.\,\footnote{Another possible way to obtain the first law of thermodynamics is through a Hamilton-Jacobi-like analysis as discussed in~\cite{Tian:2014goa}, after considering the holographic renormalization.} While we have demonstrated the first law of thermodynamic for homogeneous black holes,  it will be interesting to consider black holes that break translational invariance, in particular, for the case with an inhomogeneous background geometry. 

\section*{Acknowledgements}
I am grateful to Rong-Gen Cai, Qiang Wen, Song He and Hong Lü for helpful discussions and comments on the manuscript.
This work was supported in part by the National Key Research and Development Program of China Grant No.2020YFC2201501 and by the National Natural Science Foundation of China Grants No.12075298, No.11991052 and No.12047503.

\appendix
\section{Scalar Mass Saturating the BF Bound}\label{Sec:BF}
There is a special case for which the scalar mass $m^2$ saturates the BF bound. It is interesting and necessary to check if our discussion about the first law of thermodynamics is valid for this particular situation. As a concrete example, we consider the potential that appears in gauged supergravity in five dimensions~\cite{Cvetic:1999xp}:
\begin{equation}\label{vBF}
V(\phi)=-4(e^{-\frac{2}{\sqrt{6}}\phi}+2e^{\frac{1}{\sqrt{6}}\phi})\,.
\end{equation}
Near the AdS boundary where $\phi\rightarrow 0$, one has
\begin{equation}
V(\phi)=-12-2\phi^2+\mathcal{O}(\phi^3)\,,
\end{equation}
which corresponds to the cosmological constant $\Lambda=-6$ and the scalar mass $m^2=m_{BF}^2=-4$.  

The boundary asymptotics as $r\rightarrow\infty$ for the bulk fields $f(r), \eta(r)$ and $\phi(r)$ are given by
\begin{equation}\label{UVexpand5DBF}
\begin{split}
\phi(r)& = \frac{\phi_s\ln(r)}{r^2}+\frac{\phi_v}{r^2}+\frac{\ln(r)^2}{2\sqrt{6}r^4}\phi_s^2+\frac{\ln(r)}{\sqrt{6}r^4}\phi_s(\phi_s+\phi_v)+\frac{3\phi_s^2+4\phi_s\phi_v+2\phi_v^2}{4\sqrt{6}r^4}+\mathcal{O}(\frac{\ln(r)^2}{r^6})\,, \\
f(r)&=r^2\left[1+\frac{\ln(r)^2}{3r^4}\phi_s^2-\frac{\ln(r)}{6 r^4}\phi_s(\phi_s-4\phi_v)+\frac{f_v}{r^4}+\mathcal{O}(\frac{\ln(r)^3}{r^6})\right]\,,\\
\eta(r)&=\eta_0+\frac{\ln(r)^2}{3r^4}\phi_s^2-\frac{\ln(r)}{6 r^4}\phi_s(\phi_s-4\phi_v)+\frac{\phi_s^2-4\phi_s\phi_v+8\phi_v^2}{24 r^4}+\mathcal{O}(\frac{\ln(r)^3}{r^6})\,.
\end{split}
\end{equation}
Note that there are complicated logarithmic terms in the presence of non-trivial source $\phi_s$. We will set $\eta_0=0$ to fix the normalization of time.

To obtain the physical observables via the holographic renormalization procedure, we need to add appropriate boundary terms at the AdS boundary. 
For the potential $V$ in~\eqref{vBF}, the boundary terms are given by
\begin{eqnarray}\label{ct5DBF}
S_\partial=\frac{1}{2\kappa_N^2}\int_{r\rightarrow\infty}dx^4\sqrt{-h}\left[2K-6-\left(1-\frac{1}{2\ln(r)}+\frac{a}{\ln(r)^2}\right)\phi^2\right]\,,
\label{expansion}\end{eqnarray}
where $a$ is a free parameter which defines a renormalization scheme. Then we obtain the holographic stress tensor 
\begin{eqnarray}\label{TmunuBF}
T_{\mu\nu}=\frac{1}{2\kappa_N^2}\lim_{r\rightarrow\infty}r^2\left[2(K h_{\mu\nu}-K_{\mu\nu}-3 h_{\mu\nu})-\left(1-\frac{1}{2\ln(r)}+\frac{a}{\ln(r)^2}\right)\phi^2 h_{\mu\nu}\right]\,,
\label{expansion}\end{eqnarray}
and the condensate for the scalar operator
\begin{equation}\label{VevBF}
\left<O\right>=\frac{\delta}{\delta\phi_s}(S+S_\partial)_{on-shell}=\frac{1}{2\kappa_N^2}\lim_{r\rightarrow\infty}\sqrt{-h}\frac{\ln(r)}{r^2}\left[-n_\mu \nabla^\mu \phi-2\left(1-\frac{1}{2\ln(r)}+\frac{a}{\ln(r)^2}\right)\phi\right]\,.
\end{equation}
Inserting the boundary expansions~\eqref{UVexpand5DBF} into~\eqref{TmunuBF}, we obtain the energy density $\mathcal{E}$ and the pressure $P$
\begin{equation}\label{EPBF}
\begin{split}
2\kappa_N^2 T_{tt}=2\kappa_N^2\mathcal{E}&=-3 f_v+a \phi_s^2-\phi_s\phi_v+\phi_v^2\,,\\
2\kappa_N^2 (T_{xx}=T_{yy}=T_{zz})=2\kappa_N^2 P&=-f_v-\frac{6 a-1}{6}\phi_s^2+\frac{1}{3}\phi_s\phi_v+\frac{1}{3}\phi_v^2\,.
\end{split}
\end{equation}
The expectation value of the dual scalar operator obtained from~\eqref{VevBF} reads
\begin{equation}\label{condensateBF}
\left<O\right>=\frac{1}{2\kappa_N^2}\left(\phi_v-2 a \phi_s\right)\,.
\end{equation}

Employing the equations of motion~\eqref{Eoms} and using the asymptotical expansion~\eqref{UVexpand5DBF}, we obtain the free energy 
\begin{equation}
\begin{split}
\Omega=&\frac{1}{2\kappa_N^2}\lim_{r\rightarrow\infty}\left[2 e^{-\eta/2}r^2f-e^{-\eta/2}r^3\sqrt{f}\left(2K-6-\left(1-\frac{1}{2\ln(r)}+\frac{a}{\ln(r)^2}\right)\phi^2\right)\right]\,,\\
=&\frac{1}{2\kappa_N^2}\left(f_v+\frac{6 a-1}{6}\phi_s^2-\frac{1}{3}\phi_s\phi_v-\frac{1}{3}\phi_v^2\right)=-P\,,
\end{split}
\end{equation}
Evaluating the conserved quantity $\mathcal{Q}$~\eqref{eomQ} at the AdS boundary, we can obtain
\begin{equation}
\mathcal{Q}=-4 f_v+\frac{1}{6}(\phi_s^2-4\phi_s\phi_v+8\phi_v^2)=2\kappa_N^2(\mathcal{E}+P)\,,
\end{equation}
So we manage to obtain the expected thermodynamic relation
\begin{equation}\label{Free5DBF}
\Omega=\mathcal{E}-T\,s=-P\,.
\end{equation}
We now consider the first law of thermodynamics for the potential~\eqref{vBF}. 
As we have already shown, the horizon computation gives 
\begin{equation}
\frac{\delta H_h}{\Sigma}=\frac{1}{2\kappa_N^2} e^{-\eta(r_h)/2}f'(r_h)3 r_h^2\delta r_h= T\delta s\,,
\end{equation}
while evaluating at infinity yields
\begin{equation}
\begin{split}
\frac{\delta H_\infty}{\Sigma}&=\frac{1}{2\kappa_N^2}[-3\delta f_v-(\phi_s-2\phi_v)\delta\phi_v]\\
&=\delta\mathcal{E}+\frac{1}{2\kappa_N^2}\left(\phi_v-2 a\phi_s\right)\delta\phi_s=\delta\mathcal{E}+\left<O\right>\delta\phi_s\,,
\end{split}
\end{equation}
where we have used the expressions~\eqref{EPBF} and~\eqref{condensateBF}. Therefore, we have the same form of the first law of thermodynamics
\begin{equation}\label{firstlawBF}
d\mathcal{E}=T d s-\left<O\right>d\phi_s\,,
\end{equation}
even for the scalar mass that saturates the BF bound.
Note that the thermodynamic variables, such as $\mathcal{E}$ and $\left<O\right>$, depend on the parameter $a$ appearing in the boundary counterterms, the first law of thermodynamics has the same universal form.

\providecommand{\href}[2]{#2}\begingroup\raggedright\endgroup

\end{document}